\documentclass[prb,twocolumn,floatfix,nopacs]{revtex4-1}
\usepackage{graphicx,amsfonts,amssymb,amsmath,hyperref,enumerate,upgreek,tabularx,multirow,mathtools,tikz,bm,dsfont}
\usepackage[displaymath,mathlines]{lineno}
\usepackage{comment}

\newif\ifhyper
\hypertrue
\ifhyper
\hypersetup{
   citecolor = {green},
   colorlinks = {true}, 
   urlcolor = {blue} 
}
\fi

\newcommand{\beq}{\begin{equation}}
\newcommand{\eeq}{\end{equation}}
\newcommand{\beqa}{\begin{eqnarray}}
\newcommand{\eeqa}{\end{eqnarray}}
\newcommand{\ket} [1] {\vert #1 \rangle}

\newcommand{\subfig}[2]{%
    {#1}
    \vtop{%
  \vskip0pt
  \hbox{#2}
}}

\def\ket#1{\vert#1\rangle}

\def\Longarrow{\protect\@lra}
\def\@lra{\relbar\joinrel\relbar\joinrel\relbar\joinrel%
          \relbar\joinrel\rightarrow}

\begin{document}


\title{Learning of error statistics for the detection of quantum phases}

\author{Amit Jamadagni}
\email{amit.jamadagni@itp.uni-hannover.de}
\affiliation{Institut f\"ur Theoretische Physik, Leibniz Universit\"at Hannover, Appelstra{\ss}e 2, 30167 Hannover, Germany}
\author{Javad Kazemi}
\affiliation{Institut f\"ur Theoretische Physik, Leibniz Universit\"at Hannover, Appelstra{\ss}e 2, 30167 Hannover, Germany}
\author{Hendrik Weimer}
\affiliation{Institut f\"ur Theoretische Physik, Leibniz Universit\"at Hannover, Appelstra{\ss}e 2, 30167 Hannover, Germany}

\begin{abstract}

  We present a binary classifier based on neural networks to detect gapped 
  quantum phases. By considering the errors on top of a suitable
  reference state describing the gapped phase, we show that a neural
  network trained on the errors can capture the correlation between
  the errors and can be used to detect the phase boundaries of the
  gapped quantum phase. We demonstrate the application of the method
  for matrix product state calculations for different quantum phases
  exhibiting local symmetry-breaking order, symmetry-protected
  topological order, and intrinsic topological order.
\end{abstract}

\maketitle

\section{Introduction}
The classification of quantum phases at zero temperature is one of the
most important tasks in condensed matter physics. While phases
characterized by local order parameters can be successfully described
in terms of spontaneous symmetry breaking \cite{Landau2008}, the
discovery of topological phases without local order presents a
significant challenge in this endeavor. Here, we show that the notion
of an order parameter can be also transfered to such exotic phases, by
representing the classification of ground states of gapped quantum
phases as a binary classification problem for a neural network.


In recent years, algorithms based on machine learning have also been
found useful in the study of condensed matter physics where different
supervised and unsupervised algorithms have been employed to analyze
both classical~\cite{Carrasquilla2017, Beach2018} and quantum phase
transitions~\cite{Dong2019, Zhang2017b, Zhang2017c, Broecker2017,
  Hsin2021}, to compute the ground state
wavefunctions~\cite{Carleo2017, Deng2017b} and their
properties~\cite{Deng2017}. While efforts are in progress to apply
classical machine learning algorithms to gain insights into quantum
systems, there has also been interest in the emerging field of quantum
machine learning~\cite{Biamonte2017} where quantum variants of the
classical machine learning concepts are being proposed with
applications in condensed matter physics~\cite{Cong2019} and quantum
information~\cite{Beer2020, Beer2021}.

In this article, we introduce an approach to quantify the amount of
local or nonlocal order in a quantum state, based on a machine
learning method. Our approach is motivated by the characterization of
errors on top of an ordered reference state in the recently introduced
operational definition for topologically ordered quantum
phases~\cite{Jamadagni2020}. By employing conventional neural
networks, we propose a classification scheme resulting in an binary
classifier capable of detecting gapped quantum phases. In
Sec.~\ref{sec:ml_motive}, we briefly review the operational definition
and present possible generalizations. Furthermore, we outline the
machine learning based method capable of detecting different gapped
phases. To demonstrate the above introduced method, we apply this to
detect quantum phases with local order in Sec.~\ref{sec:ising},
symmetry-protected topological (SPT) order in
Sec.~\ref{sec:spt_order}, intrinsic topological order in
Sec.~\ref{sec:int_order} and their associated phase
transitions. Furthermore, in Sec.~\ref{sec:tdnn_other}, we discuss the
ability of a neural network trained on the errors of a one quantum
phase to detect another quantum phase with similar error
correlations. Finally, in Sec.~\ref{sec:summary}, we summarize the
results while also providing few directions that can be further
explored using our machine learning based method.

\section{Learning from the operational definition of topological order \label{sec:ml_motive}}

In the following, we introduce our method to detect gapped quantum
phases, which is inspired by the operational definition for
topological order \cite{Jamadagni2020}. We therefore briefly review
this approach and extensions to phases exhibiting local order
\cite{Jamadagni2021}. Essentially, this operational definition
interprets topological order as the intrinsic ability of a system to
perform topological error correction and classifies topological phases
in terms of reference states without any errors. An important step is
the identification of errors on top of the reference state. Following
the identification of the errors, it further requires defining an
error correction circuit which annihilates all errors, thereby
projecting back to the reference state. States are classified as
topologically ordered if the quantum state under consideration can be
corrected to the reference state by a circuit whose depth remains
finite in the thermodynamic limit, i.e., the time required to complete
the error correction does not diverge. To summarize, the operational
definition has two parts: (i) identification of an appropriate
excitation basis for the errors with respect to a reference state and
(ii) an appropriate error correction circuit leading to the
computation of the circuit depth. In the present article, we provide a
drastic simplification of the second step in the operational
definition, by removing the requirement to construct an error
correction circuits and replacing it with a machine learning approach.

The operational definition can be generalized towards phases
exhibiting symmetry-protected topological order and phases undergoing
spontaneous symmetry breaking exhibiting local order parameters, by
imposing symmetry constraints on the error correction circuits
\cite{Jamadagni2021}. In the following, we will also discuss how these
generalizations can be implemented within our machine learning
approach.

\begin{figure}[t]
\begin{center}
 \includegraphics[width=0.75\linewidth]{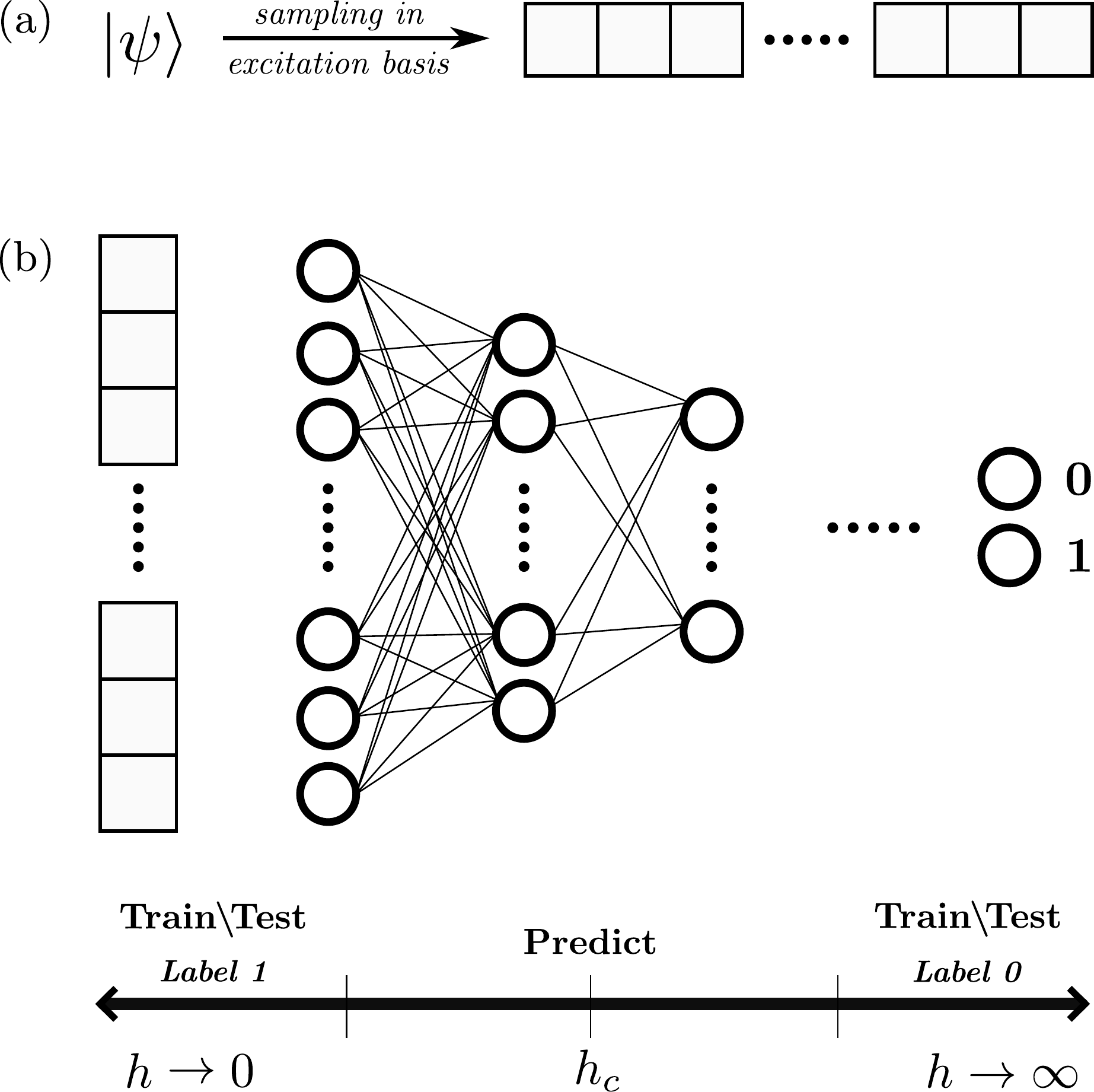}
\end{center}
\caption{Scheme for machine learning of gapped quantum phases. (a) As
  the first step, the ground state wave function is sampled in the
  error basis, yielding a classical set of measurement results for
  each sample. (b) The error sets sampled deep inside the ordered
  (trivial) phase are labeled as 1 (0) and are passed as inputs to
  train a Deep Neural Network (DNN). To extract the critical point, we
  sample the wavefunction at intermediate strengths $h$, i.e., outside
  the regime used for the training of the DNN. The prediction
  probablity for the ordered phase, $P_1$, can serve as a binary
  classifier for the phase transition.}
\label{fig:schematic_rep}
\end{figure}

The depth of the error correction circuit fundamentally captures the
deconfinement of the errors, which signals the breakdown of the
ordered phase. This deconfinement is contained in the correlations
between the errors and provides a natural way to employ machine
learning for the classification of quantum phases. Neural
network-based algorithms belonging to the class of supervised machine
learning algorithms have been effective in capturing correlations in
data and therefore are suitable candidates for the task of learning
the correlations between the errors. The key advantage of employing
such neural network-based classifiers is that they are not only
efficient at learning the correlations but also that the trained
neural network is capable of classifying errors sampled on a
wavefunction which is different from the one used for training the
neural network. In other words, given any state and the errors
obtained by sampling in the excitation basis defined with respect to a
reference state, the trained neural network is capable of predicting
whether the given state belongs to the same phase as the reference
state.

Learning the properties of the error correlations requires the
generation of suitable training data. For this, we weakly perturb the
reference state as well as a disordered paramagnetic state, for which
we know which phase the resulting states are in. Using a Monte-Carlo
sampling \cite{Jamadagni2021}, we obtain simulated measurement data
representing snapshots of the errors in the perturbed ground
state. This training set is then fed into a deep neural network (DNN),
see. Fig.~\ref{fig:schematic_rep}, which is trained using standard
machine learning algorithms for classification problems. The binary
output variable of the DNN is given by the classification into the
ordered phase connected to the reference state (1) or the disordered
paramagnet (0). The trained network is then used to predict at which
strength of the perturbation the errors become deconfined, i.e., where
the phase transition takes place. The prediction probability for the
ordered state, $P_1$, can then be interpreted as a binary classifier
for the phase boundary, as at criticality, the neural network becomes
confused \cite{vanNieuwenburg2017} and $P_1 \approx P_0 \approx 0.5$.

\section{Detecting quantum phases with local order \label{sec:ising}}
Quantum phases with local order respect the Landau symmetry breaking principle and thus the phase 
transition can be characterized by a local order parameter. One of the paradigmatic model encoding
such a phase transition is the Ising model in the presence of a transverse field. Here, we consider a spin-1/2 chain with open boundary conditions with an antiferromagnetic (AFM)
nearest neighbor interaction $(J>0)$ in the presence of a transverse field, whose Hamiltonian is 
given by
\begin{equation}
  H_{\text{TFI}} = J\sum\limits_{i}^{N-1}\sigma_{z}^{i}\sigma_{z}^{i+1} - h_{x}\sum\limits_{j}^{N}\sigma_{x}^{j}.
  \label{eq:ising}
\end{equation}
    
This model has been extensively analyzed in the literature, and it is
well known that in the limit of $J \gg h_{x}$, the ground state
exhibits antiferromagnetic order, while in the limit of $ J \ll
h_{x}$, the ground state is a paramagnet with the spins aligned in the
$x$-direction. The critical point for the phase transition is located
at $h_{x}/J = 1$. For convenience, we set $J=1$ in the following.

\subsection{Errors associated with the AFM phase}
For applying the machine learning procedure, we begin by first
introducing the errors associated to the AFM phase. To this extent, we
consider the degenerate ground state in the antiferromagnetic limit of
$h_{x}/J \rightarrow 0$ in Eq.~\ref{eq:ising} given by
$\ket{1010...10}$ or by $\ket{0101...01}$, where $0$ and $1$ denote a
spin pointing down and up, respectively. The errors in the state are
identified by the presence of domain walls which appear when two
neighbors have the same spin. As the transverse field is turned on,
the perturbed ground states exhibit virtual domain wall
excitations. Below the transition, the domain walls are bound, while
above the transition the domain walls become free and proliferate
through the entire system. Hence, the phase transition can be
understood as a confinement-deconfinement transition of the domain
walls and the correlations between the domain walls can be used to
characterize the AFM phase.

Having introduced the notion of domain walls, i.e., errors associated with the AFM phase, we now 
outline the procedure to generate errors at a finite strength of the transverse field $h_{x}$. For this, we consider the Matrix Product State (MPS) representation of the perturbed ground state computed
using the Density Matrix Renormalization Group (DMRG) algorithm using the ITensor library\cite{Fishman2020}. 
We sample the MPS computed at various strengths of the transverse field in the $\sigma_{z}$ basis, i.e., given 
an MPS we simulate the measurement outcomes using a Monte-Carlo technique as in Refs.~\onlinecite{Han2018, 
Jamadagni2021}. From the sampled data, we construct the errors by observing the nearest neighbor of a given site and
denote the presence (absence) of domain wall with 1 (0).

\subsection{Training a DNN with errors}
To recognize the correlations between the errors, we turn to supervised machine learning 
algorithms and employ a DNN. A DNN is characterized by an input layer, output layer and multiple
hidden layers with each layer consisting of multiple nodes. These nodes are further connected by 
edges which carry a weight. Here, we consider an all-to-all connected DNN i.e., 
all the nodes are interconnected. As DNNs belong to the class of supervised machine 
learning algorithms, a labeled dataset is required for the purposes of training i.e., optimizing
the variable weighted edges to map the input data to their appropriate labels. To estimate the 
performance of a DNN, we randomly split the labeled dataset into training and validation datasets, with 
the former being used for training and the latter being used to benchmark the training.

Having introduced the key ingredients of DNNs, we now outline the procedure to 
train the errors of the AFM phase at a finite field strength $h_{x}$. For our training prodcuedure, we need labeled inputs to train and validate the DNN.
For this, we sample the ground state of Eq.~\ref{eq:ising} at various $h_{x}$ deep inside 
the AFM phase, i.e., $0.25<h_{x}<0.75$, and deep in the trivial phase, i.e.,$1.5<h_{x}<2$, to construct the respective 
errors. We then label the errors in the AFM (trivial) phase with 1 (0) for the purpose of training
the DNN. For details on the structure of the DNN, a binary classifier, and other associated training 
parameters, see App.~\ref{app:A}.

\begin{figure*}[t]
\begin{center}
  \begin{tabular}{cp{0.01mm}cp{0.01mm}c}
    \subfig{(a)}{\includegraphics[height=4.75cm]{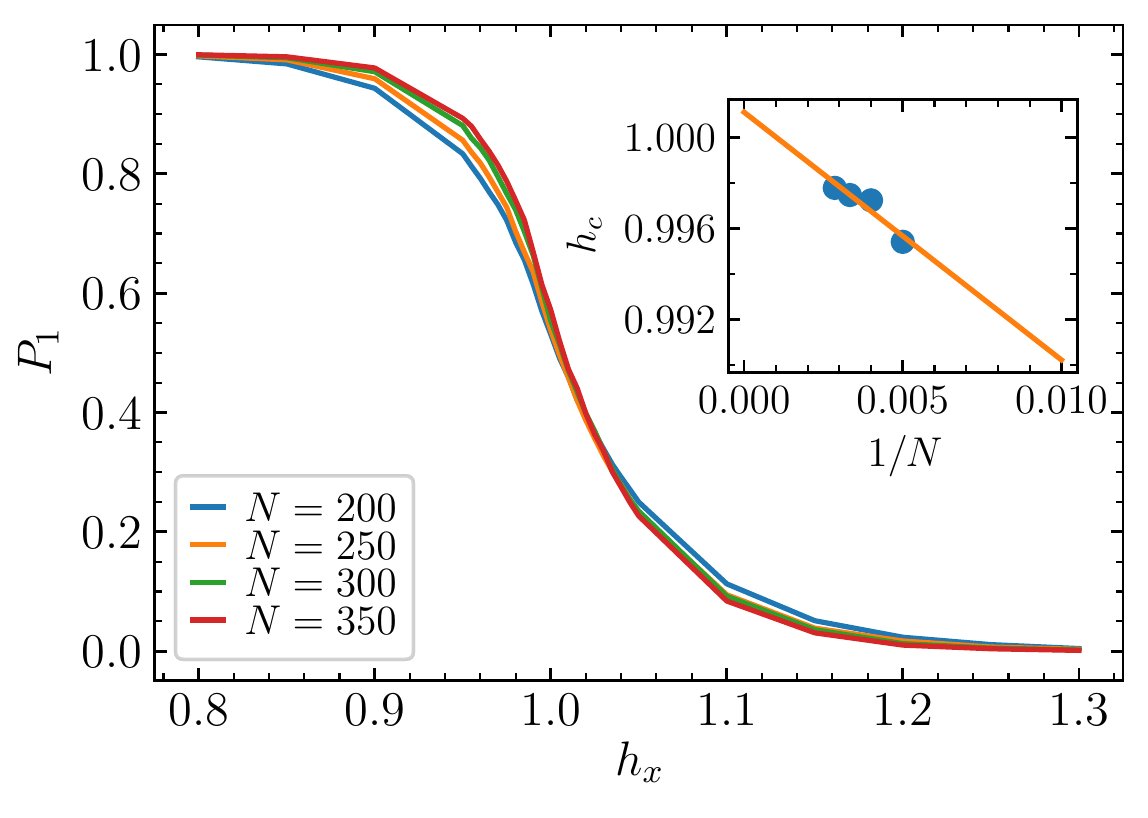}}
    &&
    \subfig{(b)}{\includegraphics[height=4.75cm]{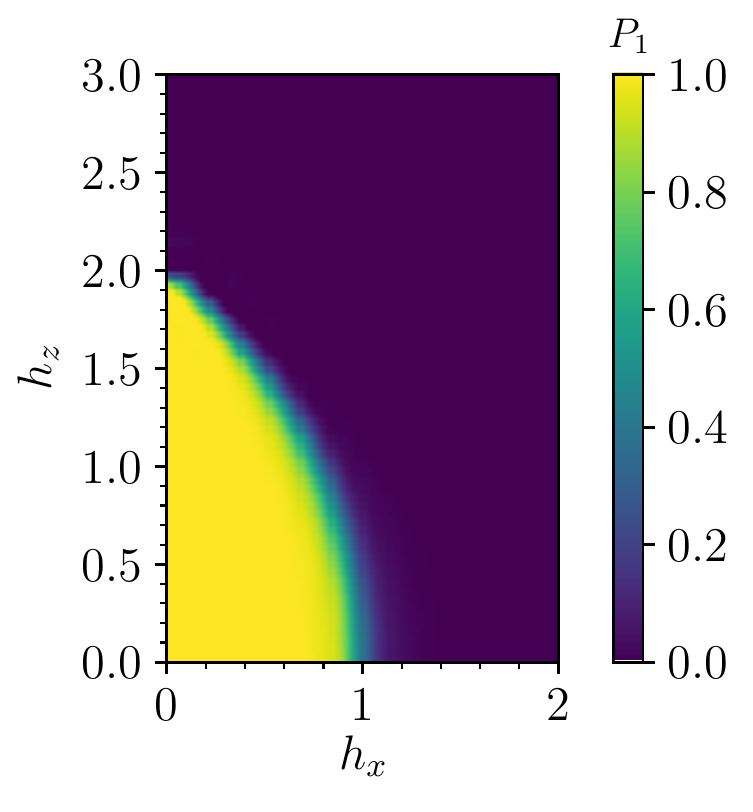}}
    &&
    \subfig{(c)}{\includegraphics[height=4.75cm]{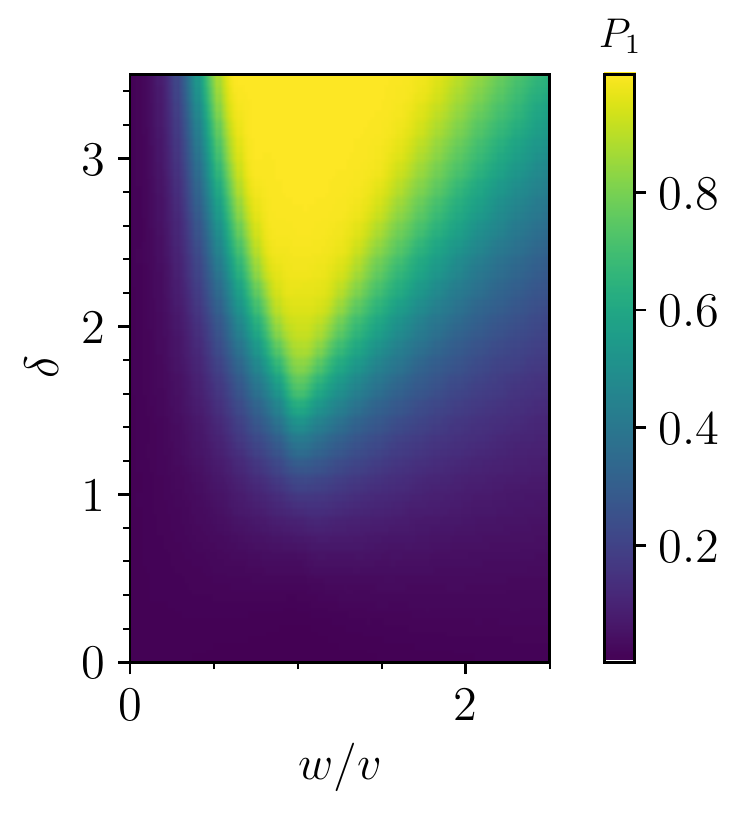}}\\
  \end{tabular}
\end{center}
\caption{Detecting antiferromagnetic order using machine learning. 
  (a) Classification probability $P_1$ for antiferromagnetic order close to the phase transition and outside of the training regime. The probability of measured errors 
  belonging to the AFM phase is 1 below the transition and approaches zero above the critical point. The inset shows the finite size scaling analysis for the critical value, leading to $h_{c}=1.001(1)$ 
(b) $P_1$ for a DNN trained on the errors of the AFM phase of the transverse field Ising model 
  is used to detect the antiferromagnetic order in the presence of a longitudinal field.  (c) The same
  DNN is used to map out the AFM phase in the Extended Bosonic SSH model,
  captured results in (b) and (c) are shown for $N=100$ spins.}
  \label{fig:ising_af}
\end{figure*}

\subsection{Detecting quantum criticality using a DNN}

In the following, we will be interested to determine the critical
point of the phase transition between the AFM and the paramagnet using
our trained DNN. For this, we sample the MPS representation of the
ground state wave function at different intermediate values of the
field strength $h_x$. The resulting errors are classified by the
trained DNN, resulting in a probability $P_1$ for the classification
in the ordered phase. Likewise, $P_0$ represents the probability for
the classification in the disordered phase. As shown in
Fig.~\ref{fig:ising_af}(a), $P_1$ is very close to one until one gets
very close to the quantum phase transition. The location of the phase
transition can be further narrowed down by a finite size scaling
analysis. The inset of Fig.~\ref{fig:ising_af}(a) shows the position
of the maximum of $\partial P_1/\partial h_x$, predicting the critical
point to be located at $h_x = 1.001(1)$, which is in excellent
agreement with the exact value of $h_x = 1$. Remarkably, 
the accuracy of the DNN on the training and the validation dataset 
is close to 100\% leading to negligible error bars in the finite size 
scaling analysis.

\subsection{Detecting antiferromagnetic order under different perturbations}

One of the significant features of the introduced machine learning
method is that the trained DNN can now be utilized to detect
antiferromagnetic order under different perturbations. To demonstrate
this, we consider two cases: (a) The Ising model in the presence of
both a transverse and a longitudinal field, (b) an extended bosonic
Su-Schrieffer–Heeger (SSH) model. Here, we detect the phase boundaries
for AFM order based on the DNN trained on the Ising model in only a
transverse field. Again, we sample the perturbed ground states to
construct the errors formed by domain wall excitations and pass the
Monte-Carlo samples into the trained DNN for binary classification in
terms of the classification probability $P_{1}$.

\subsubsection{Ising model with transverse and longitudinal field}

As the first case, we add an additional longitudinal field $h_z$ to the transverse field Ising model $H_\text{TFI}$ of Eq.~(\ref{eq:ising}), i.e.,
\begin{equation}
  H_{\text{TLFI}} = H_{\text{TFI}} - h_{z}\sum\limits_{j}^{N}\sigma_{z}^{j}.
  \label{eq:ising_lf}
\end{equation}

We sample the MPS representation of the ground state wave function
obtained using the DMRG algorithm at different strengths of $h_x$ and
$h_z$. We construct the corresponding errors from the sampled
wavefunction and provide these as input to the trained DNN to obtain
the classification probability $P_1$ for AFM ordering. The phase
diagram computed within our machine learning method is shown in
Fig.~\ref{fig:ising_af}(b), which is in strong agreement with the
previously established phase diagram using a local order parameter
\cite{Ovchinnikov2003}. The trained DNN is capable of not only
detecting the second order phase transition at $h_z \neq 0$ and $h_x
\neq 0$ but also detects the first order phase transition phase
transition at $h_z \neq 0$ and $h_x = 0$. Extracting the order of the
phase transition using the machine learning remains an open task for
future work.

\subsubsection{Extended bosonic SSH model}

The Hamiltonian of the extended bosonic SSH model~\cite{Elben2020} is given by
\begin{equation}
 \begin{split}
  H_{\text{BSSH}} &= \frac{v}{2}\sum\limits_{i=1}^{N/2}\big(\sigma^{2i-1}_{x}\sigma^{2i}_{x} + \sigma^{2i-1}_{y}\sigma^{2i}_{y} + \delta\sigma^{2i-1}_{z}\sigma^{2i}_{z}\big) \\
         &+ \frac{w}{2}\sum\limits_{i=1}^{N/2-1}\big(\sigma^{2i}_{x}\sigma^{2i+1}_{x} + \sigma^{2i}_{y}\sigma^{2i+1}_{y} + \delta\sigma^{2i}_{z}\sigma^{2i+1}_{z}\big).
 \end{split}
 \label{eq:ebssh}
\end{equation}
While the model is mostly interesting for its topological properties,
see below, it also features an AFM phase for large values of
$\delta$. As in the earlier case, we sample the MPS representation of
the ground state wavefunction at various strengths of the tuple $(w/v,
\delta)$. The errors associated with the AFM phase are passed as
inputs to the trained DNN output $P_{1}$ as before.  The
antiferromagnetic order computed using the DNN in
Fig.~\ref{fig:ising_af}(c) is in strong agreement with the phase
diagram established using other methods~\cite{Elben2020,
  Jamadagni2021}. Hence, we have demonstrated that a DNN trained with
a certain perturbation of the reference state is capable of detecting
the gapped phase in the presence of different perturbations.

While it is in general straightforward to work with local order
parameters for phases exhibiting spontaneous symmetry breaking,
another key feature of our approach is that it can be extended to
gapped quantum systems with topological order, which we shall explore
in the next section.

\section{Detecting quantum phases with topological order}

Let us now establish that the machine learning method is also capable
of detecting topological phases of matter, which are beyond the
conventional Landau symmetry breaking principle. Many different
topological states of matter have been discovered \cite{Wen2019} and
their classification remains an important problem. In the following,
we limit our discussion to models exhibiting symmetry-protected
topological (SPT) order and intrinsic topological order. In the
following, we briefly review the notion of topological order based on
entanglement properties of the respective phases captured via the
local unitaries connecting them to a product state \cite{Chen2010}. In
this notion, one considers the properties of a quantum circuit
comprised of a sequence of quasi-local unitary operations. States
having topological order are said to be long-range entangled, i.e.,
any circuit that takes a topologically ordered state to a product
state has to have a circuit depth that diverges in the thermodynamic
limit.

\subsection{Detecting symmetry-protected topological order \label{sec:spt_order}}

Importantly, not all short-range entangled states that can be connected
to a product state by a finite-depth quantum circuit are topologically
trivial. If one imposes symmetries on the quantum circuits, it is
possible that a short-range entangled state can only be connected to a
product state by symmetry-preserving circuits whose depth diverges in
the thermodynamic limit. These states are said to have
symmetry-protected topological (SPT) order. In the context of the
recently introduced operational definition~\cite{Jamadagni2021}, a
state is said to be SPT ordered if it can be corrected by a symmetry
preserving finite-depth error correction circuit. In the following, we
introduce the machine learning based order parameter to detect SPT
phases in the variants of the SSH model.

\subsubsection{Phase transitions in the SSH model}
\begin{figure}
\begin{center}
 \includegraphics[width=0.75\linewidth]{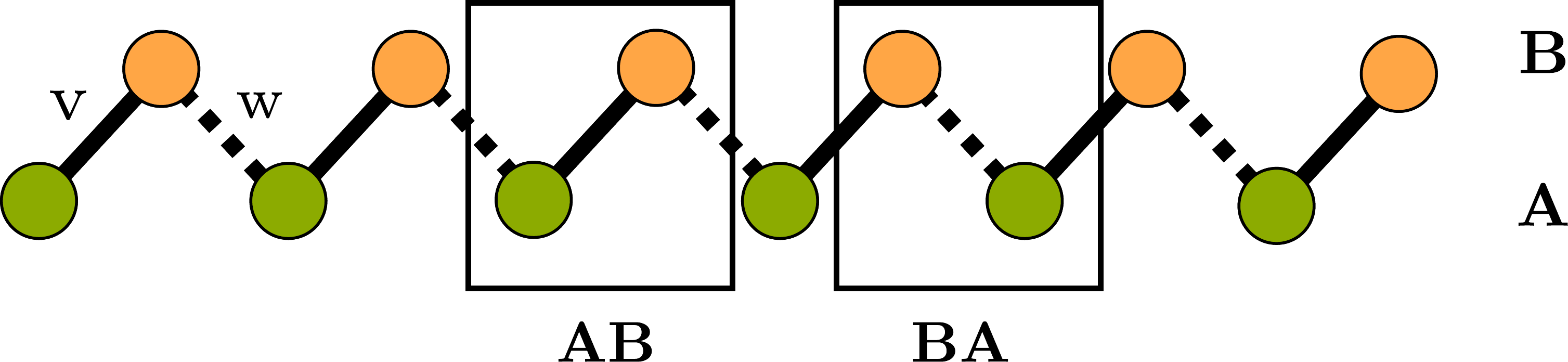}
\end{center}
\caption{The SSH Hamiltonian describes the hopping of particles on a 1D lattice with the choice 
of the unit cell as AB/BA and hopping strengths between the rails $A$ and $B$, given by $v$ and 
$w$, respectively.}
\label{fig:ssh_1d}
\end{figure}

One of the well known models exhibiting SPT order is the SSH model~\cite{Su1980}, whose Hamiltonian 
describes the hopping of a particle on a one-dimensional lattice, see Fig.~\ref{fig:ssh_1d}. We consider a hardcore boson variant, which is given by the $\delta=0$ case of Eq.~(\ref{eq:ebssh}), i.e.,
\begin{equation}
H_{\text{SSH}} = v\sum\limits_{i=1}^{N/2}\sigma_{-}^{2i-1}\sigma_{+}^{2i}
+ w\sum\limits_{i=1}^{N/2-1}\sigma_{-}^{2i}\sigma_{+}^{2i+1} + \text{h.c.}.
\label{eq:ssh}
\end{equation}

\begin{figure*}[t!]
\begin{center}
  \begin{tabular}{cp{0.01mm}cp{0.01mm}c}
    \subfig{(a)}{\includegraphics[height=4.75cm]{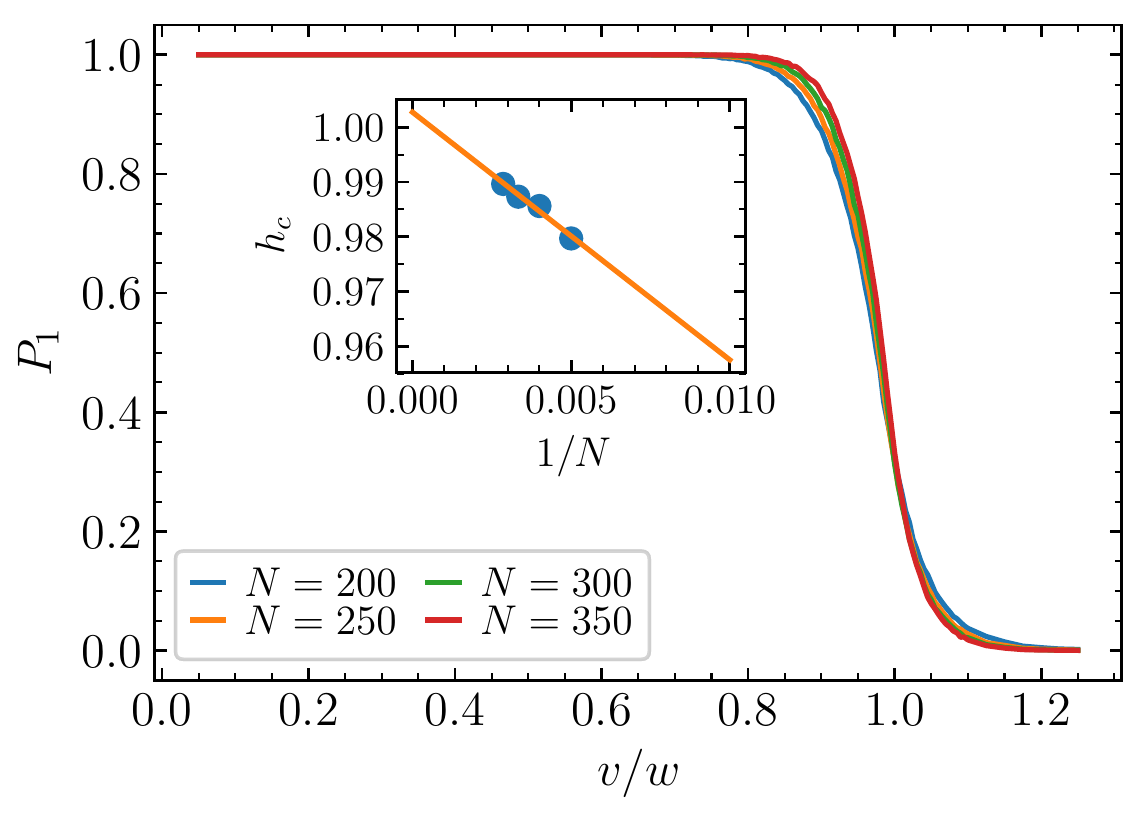}}
    &&
    \subfig{(b)}{\includegraphics[height=4.75cm]{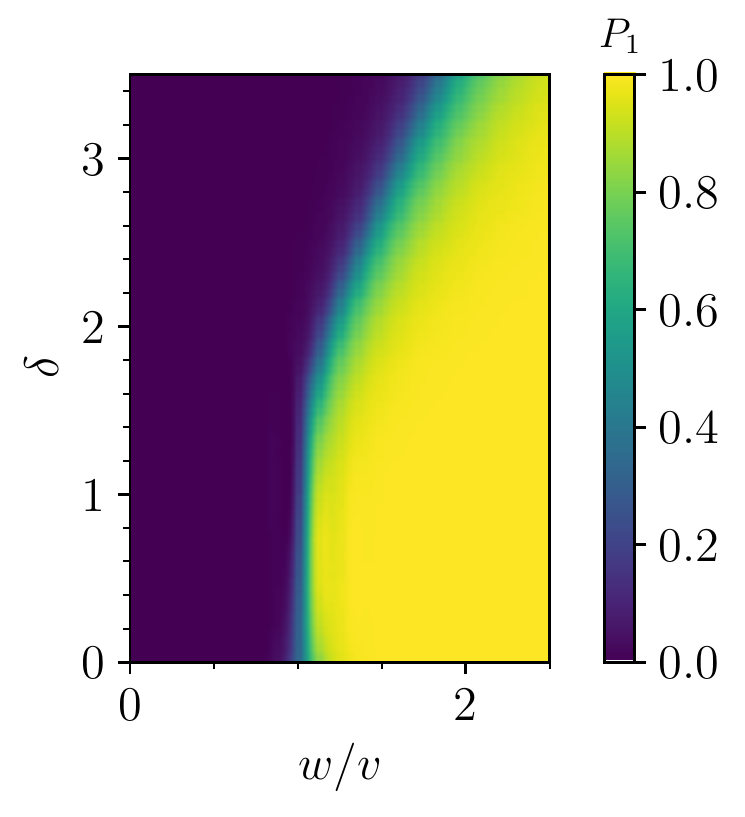}}
    &&
    \subfig{(c)}{\includegraphics[height=4.75cm]{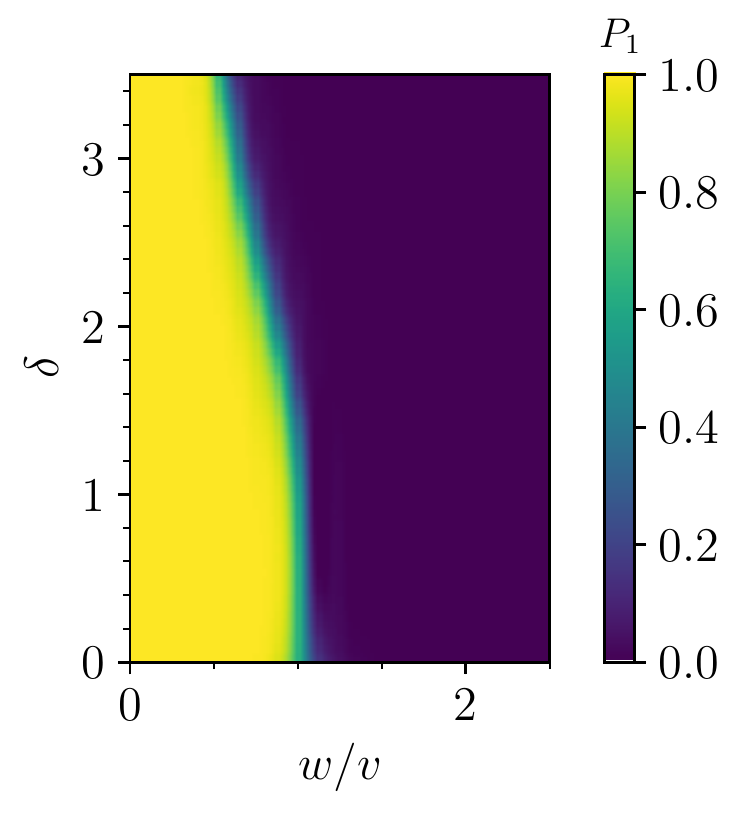}}\\
  \end{tabular}
\end{center}
\caption{Probing SPT phases in the SSH model and the extended bosonic SSH model. (a) Classification probability $P_{1}$ for the $v<w$ SPT phase as a function of the hopping strength ratio, $v/w$, 
for the choice of the unit cell as $AB$. 
The inset shows the finite size scaling analysis for the critical value, leading to $v_c=1.003(2) w$. (b) Using the DNN
trained on the errors sampled with respect to the reference state $\ket{\psi}_{\text{AB}}$  of the SSH model, we predict 
the corresponding SPT phase in the extended bosonic SSH model by computing $P_{1}$ as a function of the hopping strength ratio
$w/v$ and AFM interaction strength $\delta$. (c) Using the DNN trained on the errors on top of the reference state $\ket{\psi}_{\text{BA}}$, we map out the extent of the $v>w$ SPT phase. A system size of $N=100$ is used for computing the phase 
diagram in (b) and (c).}
\label{fig:ssh_spt}
\end{figure*}

The SSH Hamiltonian has been extensively studied, in the case of periodic boundary conditions where the Hamiltonian is exactly solvable
with gapped phases in both the limits of $v<w$ and $v>w$ with a gap closing at $v=w$ signaling a phase transition. Additionally,
the limit of $v<w$ and $v>w$ have non-vanishing winding number, $\nu=1$, which  is a characteristic feature of the topological phase.
In the case of open boundaries, in the limit of $v \ll w$ we have edge modes at the end of the chain which characterize topological
order with topological phase transition at $v=w$. Within the operational definition, one can show that the phase in the case $v>w$ is also SPT ordered, which is distinct from the SPT ordered phase characterized by edge
modes \cite{Jamadagni2021}.

Having briefly reviewed the key features of the SSH model, we turn to
the machine learning method introduced earlier to train a DNN to probe
SPT phases and their respective phase transitions. To train the DNN,
we need to generate the errors associated with a SPT phase. For this,
we briefly review the notion of errors related to the SPT phase as
introduced in the Ref.~\onlinecite{Jamadagni2021}. For both SPT
phases, the reference states are found by setting either $w=0$ or $v=0$. Then, the ground state can be found by putting singlets on the $AB$ ($BA$) unit cells, i.e.,
\begin{equation}
  \ket{\psi}_{\text{AB/BA}} = \frac{1}{\sqrt{2}}\prod\limits_{i\in B/A} (\ket{0}_i\ket{1}_{i+1}-\ket{1}_i\ket{0}_{i+1}).
  \label{eq:ssh_1d}
\end{equation}
We can introduce an excitation basis for each unit cell, consisting of
the error-free state $\ket{\bm{-}} = (\ket{01} - \ket{10})/\sqrt{2}$
as well as density fluctuations $\ket{\bm{0}} = \ket{00}$ and
$\ket{\bm{1}} = \ket{11}$, and phase fluctuations $\ket{\bm{+}} =
(\ket{01} + \ket{10})/\sqrt{2}$. In the following, we discuss the case
of the unit cell choice being $AB$, leading to SPT ordered phase in
the limit of $v<w$ which is characterized by the presence of edge
modes.

The errors in the system correspond to the density fluctuations
$\ket{\bm{0}}, \ket{\bm{1}}$ and the phase fluctuation
$\ket{\bm{+}}$. We construct the errors by sampling the MPS
representation of the ground state wavefunction in the excitation
basis at finite $v$. For the sake of training the DNN, we label the
errors $\bm{0}, \bm{1}, \bm{+}, \bm{-}$ as $\{0, 1, 2, -1\}$, with the
outputs generated in the limit of $v<w$ ($v>w$) labeled as 1 (0). As
earlier, $P_{1}$ i.e., the probability of errors being labeled as 1
captures the phase transition.  By performing finite size scaling
analysis, we extract the critical value for $v$, see Fig.~\ref{fig:ssh_spt}, and note it to be in
good agreement with the exact result $v=w$. 

The discussion on choosing the other unit cell i.e., the $BA$ unit
cell, leading to the detection of the $v>w$ SPT is similar.  The key
difference between the two scenarios is that, in the previous limit
for training purposes we label the errors in the limit of $v<w$
($v>w$) to be 1 (0) while in the current scenario we invert the above
by labeling the errors in the limit of $v>w$ ($v<w$) to be
1 (0). Training and constructing the DNN, as well as extracting the
critical point can be performed in exactly the same way.

\subsubsection{Detecting SPT order in the Extended Bosonic SSH model}

Let us now turn to the extended bosonic SSH model of
Eq.~\ref{eq:ebssh} with nonzero antiferromagnetic interaction
$\delta$. In particular, we are interested in the fate of the two SPT
phases upon increasing $\delta$, using the trained DNNs from the previous section, which encodes the error correlations
of the SPT phases for $\delta =0$ and  $v<w$ and $v>w$, respectively. To this extent, we consider the MPS representation of the
ground state of the Hamiltonian as a function of the tuple $(w/v, \delta)$. We then sample the 
MPS representation in the excitation basis defined with respect to the reference state to generate the errors 
associated with the SPT phases in the limit of $v<w$ and $v>w$ as introduced in Eq.~(\ref{eq:ssh_1d}). 
In Fig.~\ref{fig:ssh_spt}(b,~c), we plot $P_{1}$, the probability of sampled errors being labeled as 1, in the $v<w$ ($v>w$) 
regime thereby detecting the SPT phase corresponding to $v<w$ ($v>w$) of the Extended Bosonic SSH Hamiltonian. The phase 
diagram of the extended bosonic SSH model computed using the neural network-based method is in very good agreement with other methods~\cite{Elben2020, Jamadagni2020}.

\subsection{Intrinsic Topological order \label{sec:int_order}}

In contrast to symmetry-protected topological order, intrinsic topological order refers to states that cannot be mapped onto product states by a finite-depth quantum circuit, even if arbitrary quasi-local unitaries are allowed. In the context of the operational definition, topological order is quantified
by the error correcting abilities of a state. Formally, the operational definition quantifies a state to be topologically 
ordered if it can be corrected by a finite-depth error correction circuit~\cite{Jamadagni2020}. One of the paradigmatic 
models exhibiting intrinsic topological order is the toric code model~\cite{Kitaev2003}. In the following, we 
briefly introduce variants of the toric code model and discuss the robustness of topological order in various perturbed 
toric code models using the machine learning method.

\begin{figure*}
\begin{center}
  \begin{tabular}{cp{0.01mm}cp{0.01mm}c}
    \subfig{(a)}{\includegraphics[height=4cm]{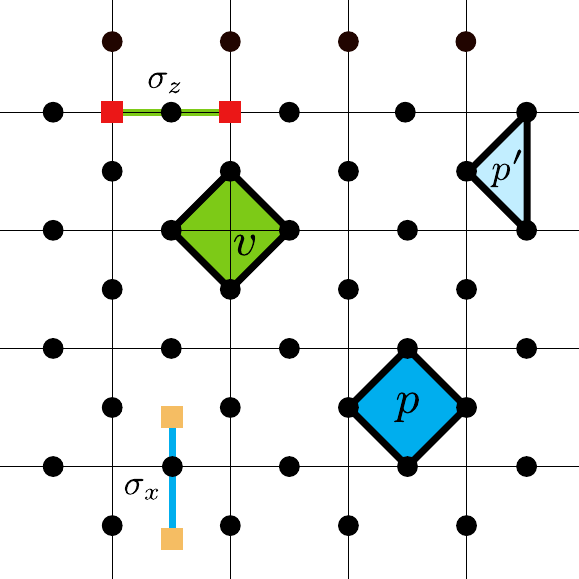}}
    &&
    \subfig{(b)}{\includegraphics[height=4cm]{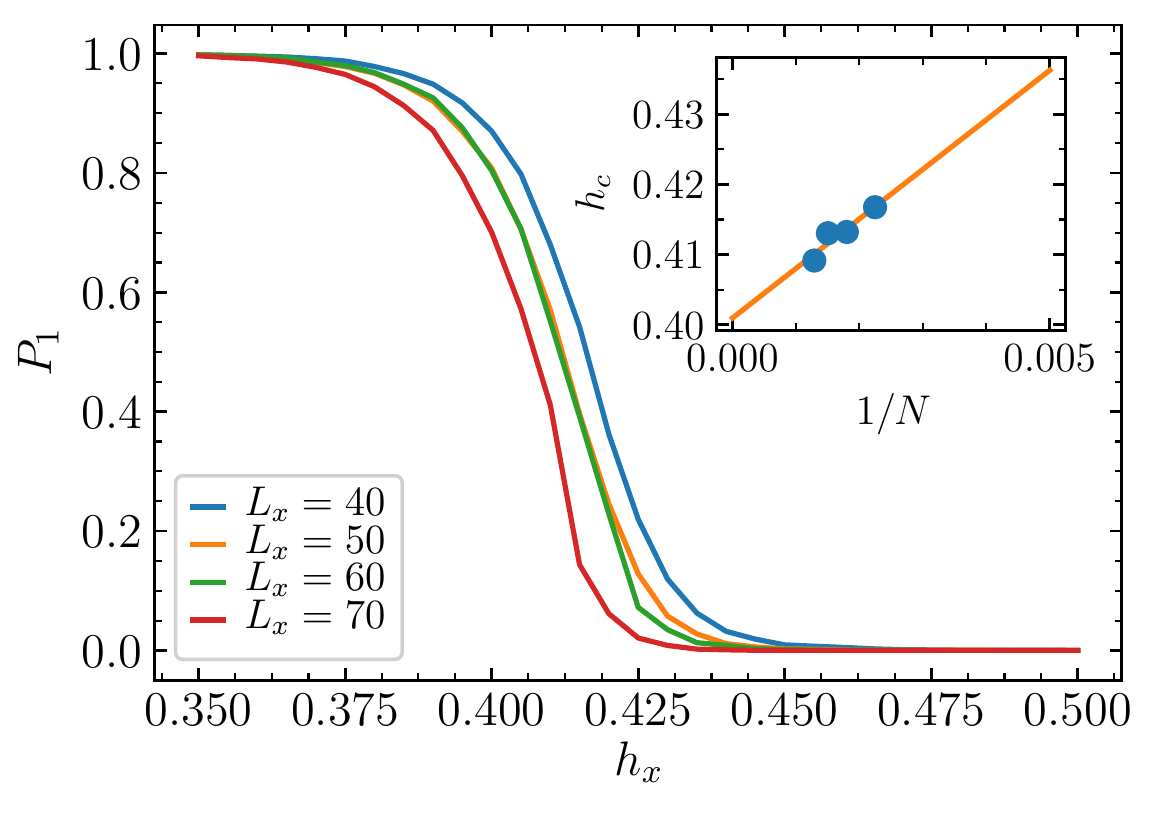}}
    &&
    \subfig{(c)}{\includegraphics[height=4cm]{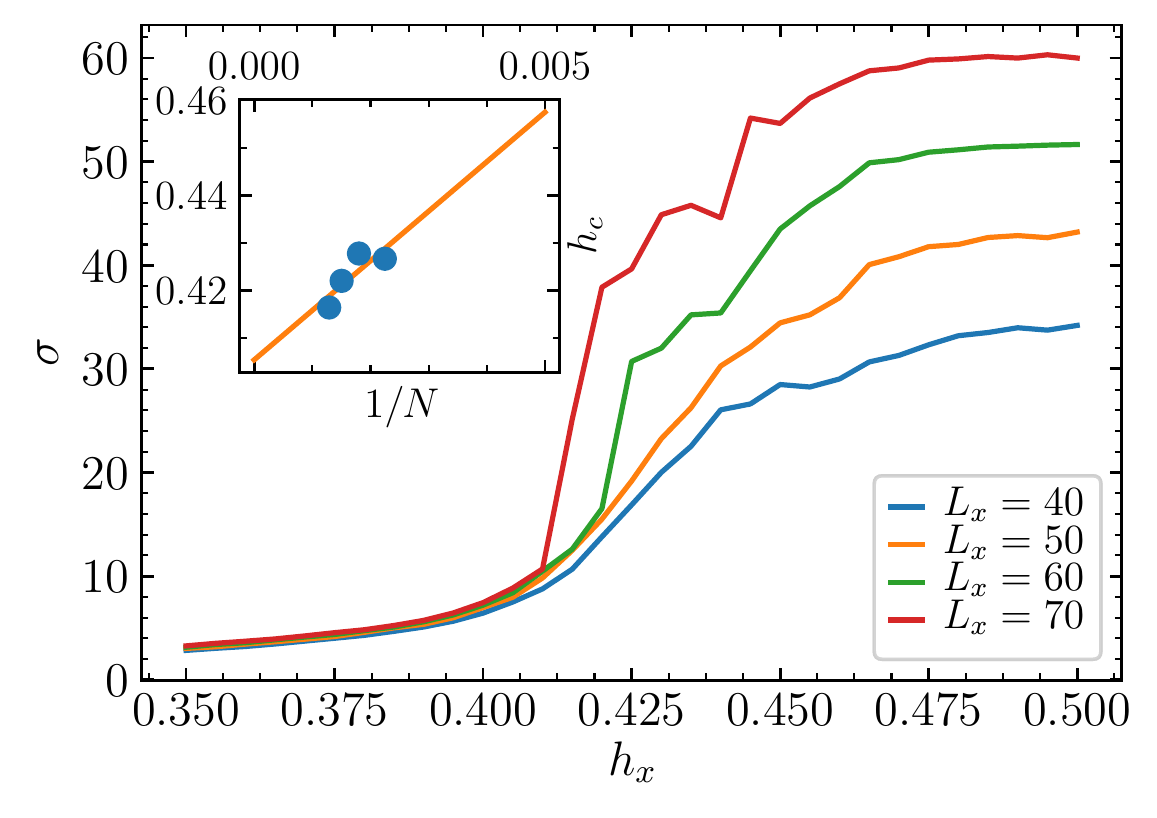}}\\
  \end{tabular}
\end{center}
\caption{(a) Toric code with rough open boundaries with the vertex and face operators given by $A_{v}$ (green diamond), 
$B_{p}$ (blue diamond) and $B_{p'}$ (light blue diamond). The anticommutators of $A_{v}$ ($B_{p/p'}$) 
  given by $\sigma_{z(x)}$ generate the vertex (face) excitations denoted by red (orange) squares. (b) Classification probability $P_{1}$ for topological order, as predicted by the trained DNN, as a function of the external perturbation strength, $h_{x}$. The inset shows the phase transition between the topologically ordered phase and the trivial paramagnet to occur at $h_{c}^{ml}=0.401(3)$. (c) Standard deviation 
$\sigma$ of the error correction circuit depth according to the operational definition for topological order, resulting in a critical strength of $h_{c}^{od}=0.405(8)$ (inset). Results are shown for a
$L_{y} \times L_{x}$ lattice, where $L_{y(x)}$ is the number of rows (columns). We fix $L_{y}=5$ while 
varying $L_{x}$, with the total number of spins, $N$, given by $(2L_y + 1)\times L_x + L_y$.}
\label{fig:toric_hx}
\end{figure*}

\subsubsection{Phase transitions in the toric code model}
To introduce the toric code model, we consider a square lattice with open boundary conditions with spins on the edges as in 
Fig.~\ref{fig:toric_hx}. The Hamiltonian is given by
\begin{equation}
H_{\text{TC}} = -\sum\limits_v A_{v} - \sum\limits_p B_{p} - \sum\limits_{p'} B_{p'},
\label{eq:tc}
\end{equation}
where $A_{v} = \prod_i\sigma_{x}^{i}$, $B_{p} =
\prod_j\sigma_{z}^{j}$, for $i, j$ being the spins attached to the
vertex $v$ and face $p$ respectively. The open boundaries in terms of
$B_{p'}$ operators, which realizes a three-body $\sigma_{z}$
interaction, in contrast of the four-body interaction of $B_p$. Such
boundary conditions are generally referred to as rough boundaries, for
a more detailed description on various boundary conditions associated
with the toric code, we refer the reader to
Ref.~\onlinecite{Beigi2011}. The ground state of the toric code
Hamiltonian on a planar geometry with rough boundaries is given by
$\mathcal{N}\prod\limits_v(\mathds{1} + A_{v})\ket{\textbf{0}}$, where
$\ket{\textbf{0}} = \ket{000...0}$. The excitations in the system are
generated by the action of the anticommutators of $A_{v}$ ($B_{p}$)
given by $\sigma_{z(x)}$ on the ground state and are identified as
vertex (face) excitations.

To demonstrate the machine learning based method, we consider the
toric code with rough boundaries in the presence of an external
magnetic field pointing in the $x$ direction, leading to the
Hamiltonian, $H_{\text{PTC}} = H_{\text{TC}} -
h_{x}\sum_{i}\sigma_{x}^{i}$.  In the limit of $h_{x}\rightarrow0$,
the phase is topologically ordered while in the limit of
$h_{x}\rightarrow\infty$ we have a paramagnetic phase. To capture the
phase transition using the DNN, we compute the MPS representation of
the ground state wavefunction as a function of the perturbation
strength $h_{x}$, using a DMRG algorithm. We then sample the
wavefunction in the $\sigma_{z}$ basis to construct the
errors. Observing the fact the perturbation only anticommutes with the
$B_{p}$ operator, we conclude that the system only hosts face
excitations. To capture the face excitations and therefore to
construct the errors, it is sufficient to compute the parity of ones
in each face by measuring each spin in the $\sigma_{z}$ basis. Even
(odd) parity of ones indicates the absence (presence) of excitation
leading to construction of the errors. As established earlier, we
label the errors in the topological (trivial) phase as 1 (0) to train
the DNN. The trained DNN is now exposed to errors sampled outside of
the training regime and $P_{1}$, the probability of errors being
labeled as 1, capture the phase transition, with the critical strength
being computed by performing finite size scaling analysis, as in
Fig.~\ref{fig:toric_hx}. To validate the transition point obtained by
the machine learning method, we compare it with the critical value
obtained from the operational definition, see
Fig.~\ref{fig:toric_hx}. We note the values for the critical
perturbation strength obtained using both methods are in good
agreement and therefore further validate the machine learning based
method. Furthermore, we would like to point out that, for the same
number of samples, $P_1$ obtained from machine learning exhibits
significantly less noise than the standard deviation of the circuit
depth $\sigma$ within the operational definition, i.e., the machine
learning approach requires less computational resources for the same
level of accuracy. For the details on the error correction circuit
used to extract the time statistics, see App.~\ref{app:B}.

\section{Detecting other gapped phases with similar errors \label{sec:tdnn_other}}

In the previous sections, we have established that a DNN trained on the errors of a gapped quantum phase is capable of 
detecting the same phase even if the perturbation is changed. As introduced in Sec.~\ref{sec:ml_motive}, a gapped 
quantum phase has a correspondence with the associated errors and their correlations. However, a DNN trained
on the error correlations of a gapped phase is still capable of recognizing other gapped quantum phases with similar 
error correlations. For instance, consider gapped phases $A$ and $B$ whose error correlations \textit{almost} remain 
the same, this implies that a DNN trained on the error correlations of $A$ is capable of identifying the gapped phase 
$B$. However, the error correlations remaining \textit{exactly} the same does not guarantee the equivalence of the 
gapped quantum phases as the errors themselves might be different. In this section, we discuss two examples: (a) detecting 
a ferromagnetic phase using a DNN trained on the errors of an antiferromagnetic phase and (b) detecting topological order in a punctured toric 
code under a perturbation using a DNN trained on the errors of an nonpunctured toric code.

\begin{figure}[b]
    \includegraphics[width=0.8\linewidth]{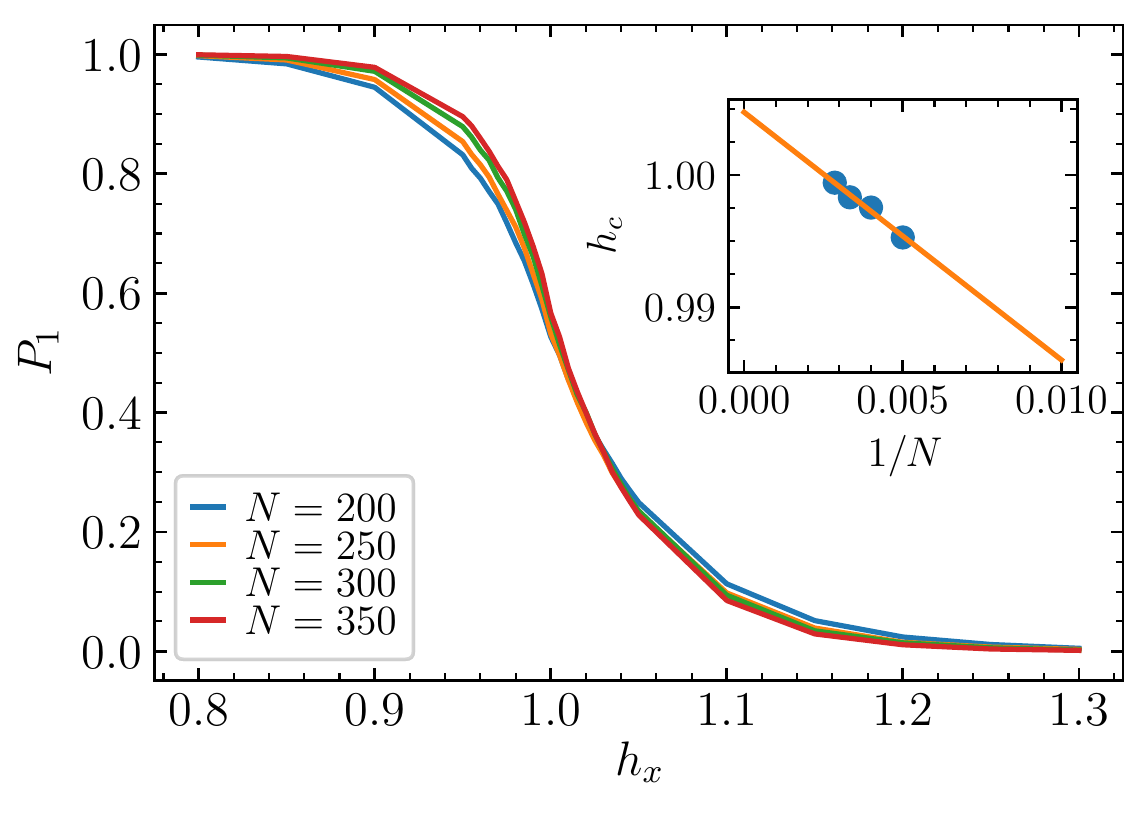}
\caption{Detecting ferromagnetic order in a transverse field Ising model with ferromagnetic interactions using 
  a DNN trained on the errors of the AFM phase, shown by the classification probability $P_1$ for ferromagnetic order as a function of the transverse field strength $h_x$. Finite size scaling results in a critical field strength of  $h_{x}^{c}=1.005(1)$ (inset).}
\label{fig:ferro_ising}
\end{figure}


\begin{figure*}[t]
\begin{center}
  \begin{tabular}{cp{0.01mm}c}
    \subfig{(a)}{\includegraphics[height=4cm]{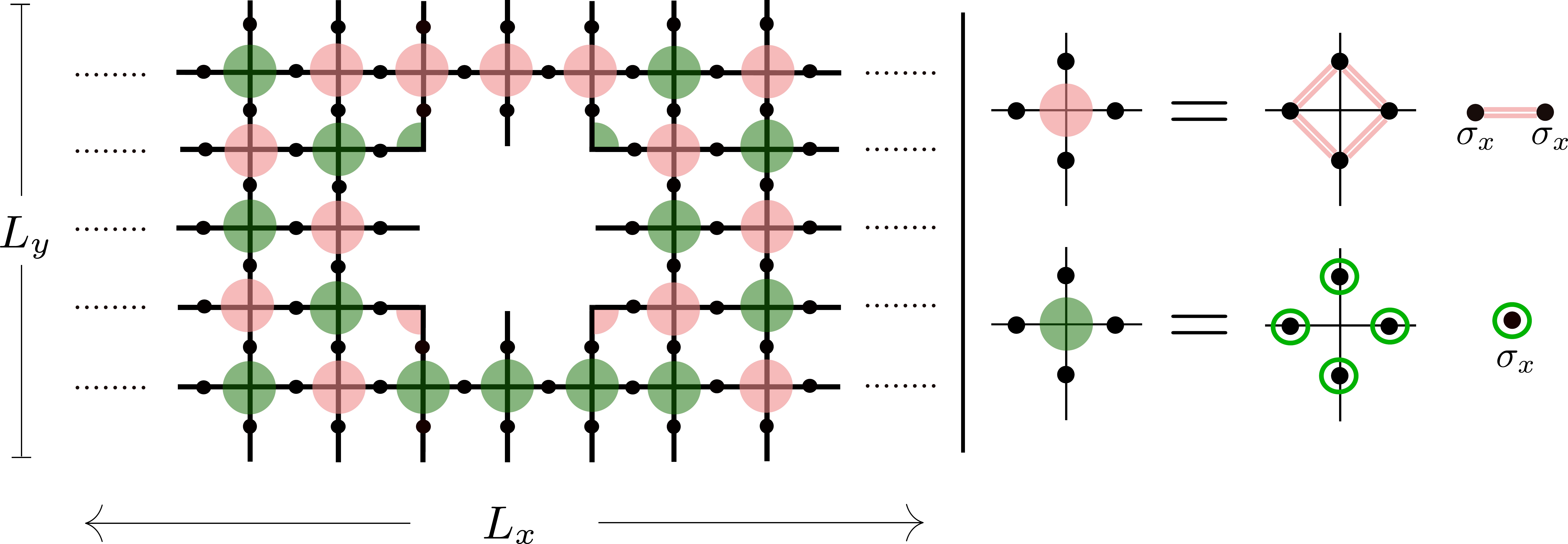}}
    &&
    \subfig{(b)}{\includegraphics[height=4cm]{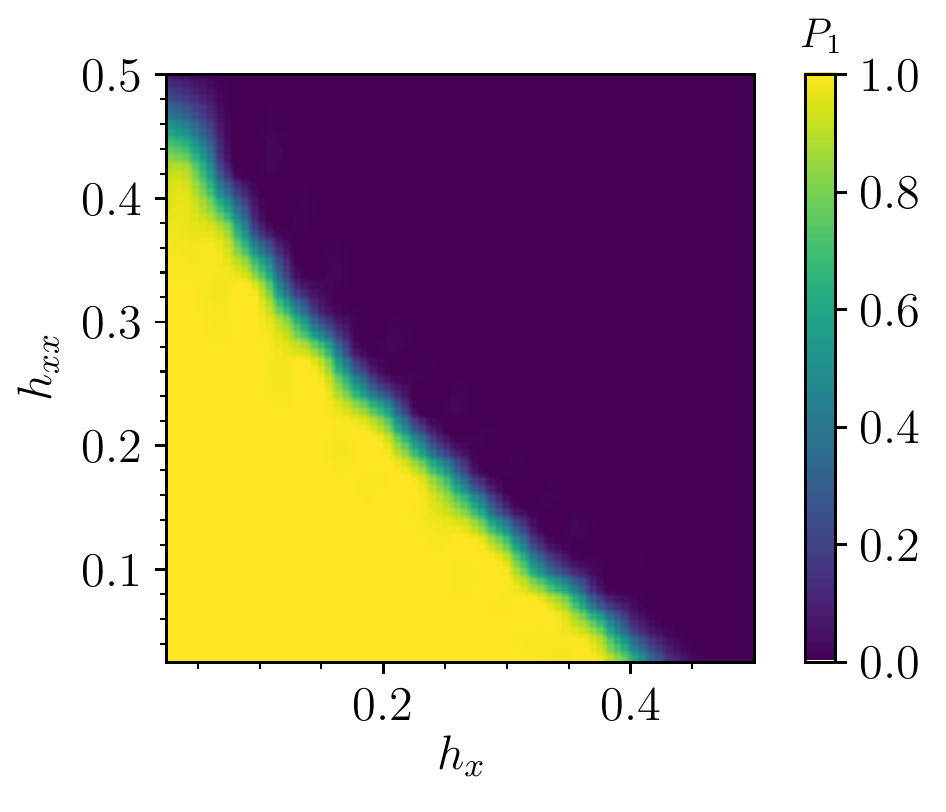}}\\
  \end{tabular}
\end{center}
\caption{(a) Punctured toric code with alternating perturbations
  according to local terms $h_x \sigma_x^i$ and two-body interactions
  $h_{xx}\sigma_x^i\sigma_x^j$. (b) The robustness of the topological
  phase of a punctured toric code as determined by the classification
  probability $P_1$ for the topologically ordered phase. The phase
  diagram is mapped out for a square lattice of $L_y \times L_x = 5
  \times 50$ with four $B_p$ interactions at the center of the lattice
  being turned off. This results in rough boundary conditions
  appearing around the puncture analogous to the $B_{p'}$ operators on
  the boundaries.}
  \label{fig:toric_hx_hxx_punc}
\end{figure*}

\subsection{Detecting ferromagnetic order using DNN trained on AFM phase}
In this section, we consider the 1D transverse field Ising model with ferromagnetic nearest neighbor interactions with 
open boundary conditions i.e., $H_{\text{TFIM}}$ as in Eq.~\ref{eq:ising} with $J<0$.


In the limit of the transverse field strength, $h_{x}\rightarrow 0$ we have a ferromagnetic phase while in the 
the limit of $h_{x}\rightarrow\infty$ we have a paramagnetic phase. The errors corresponding to the ferromagnetic phase
are given by observing the neighbors, i.e., neighbors with different (same) parity indicate the presence (absence) of 
a domain-wall (error). While the errors of the ferromagnetic phase are different than in the antiferromagnetic case, their correlations are identical as the ferromagnetic Ising model 
can be mapped onto the antiferromagnetic one by a unitary transformation flipping every second spin. In Fig.~\ref{fig:ferro_ising}, we confirm the expectation that the transition 
between the ferromagnet and the paramagnet can be detected by a DNN trained on the antiferromagnetic case.

\subsection{Detecting intrinsic topological order in perturbed punctured toric code}

In this section, we consider a case where the two models are not
connected by a unitary transformation. Here, we start from a DNN
trained on the errors of a perturbed planar toric code with rough
boundaries and apply it to a punctured toric code containing a hole in
the center, see Fig.~\ref{fig:toric_hx_hxx_punc}a. Additionally, on every second plaquette, we change the perturbation 
from a local magnetic field $h_x\sigma_x^i$ to a ferromagnetic Ising interaction $h_{xx} \sigma_x^i\sigma_x^j$. A puncture in the toric code is realized by 
turning off either the $B_{p}$ or $A_{v}$ interactions. In the current scenario, we turn off the $B_{p}$ interactions over
a small region resulting in a puncture. The topological phase 
corresponding to the punctured toric code is different from that of the nonpunctured toric code as the former has a degenerate 
ground state manifold while the latter is nondegenerate. However, it is still possible to map out the phase diagram of the 
punctured toric code in the presence of perturbation by using a DNN trained on the errors of the toric code with no punctures, 
as shown in Fig.~\ref{fig:toric_hx_hxx_punc}. For the purpose of predicting the phase diagram of the perturbed punctured toric code, we 
consider that the punctured faces do not host any errors. However, it is important to note that the above technique of estimating 
the phase diagram of the punctured toric code with a DNN trained on errors of nonpunctured toric code is appropriate only in 
the limit where few (in comparison to the total system size) interactions are turned off.

\section{Summary and Discussion \label{sec:summary}}
In summary, we have introduced a novel machine learning based
classifier to detect gapped quantum phases. Our method enhances the
operational definition to not only detect gapped quantum phases with
topological order but also other quantum phases with local
order. Furthermore, within the machine learning approach, it is no
longer necessary to construct appropriate error correction algorithms,
as this task is effectively carried out by a deep neural network
(DNN). Our work establishes a correspondence between gapped quantum
phases and their errors along with the correlations between them. In
other words, any gapped quantum phase can be uniquely identified and
quantified by its errors on top of a suitable reference state and the
correlations between them. Crucially, the DNN trained with a certain
perturbation can also successfully detect the quantum phase and its
boundaries in the presence of different perturbations.

Future avenues of our machine learning approach include (a) exploring
quantum phases in an open quantum system with a DNN trained on the
errors in a closed quantum system, (b) exploring other DNN training
routines like using random unitaries coupled with random measurements
and/or random dissipative channels, (c) quantifying the time evolution
of a quantum system out of equilibrium using machine learning
classifiers. As a further development of the method, autoencoders
could potentially be used to deduce the excitation basis and therefore
the errors associated with a given gapped quantum phase. Hence,
autoenconders in combination with DNNs could lead to a powerful
framework for the classification of any gapped phase in the framework
of machine learning. Furthermore, recent technological progress has
enabled the possible realization of topological states of matter in
quantum simulation architectures \cite{Semeghini2021,Satzinger2021},
but unambigious identification of topological order remains
challenging because of the exponential resources required to measure
quantities like the topological entanglement entropy
\cite{Preskill2006,Levin2006}. In this context, our machine learning
approach can be used to verify the existence of topological order in
these experiments in an efficient way.

\begin{acknowledgments}

  This work was funded by the Volkswagen Foundation, by the Quantum
  Valley Lower Saxony (QVLS) through the Volkswagen Foundation and the
  ministry for science and culture of Lower Saxony, by the Deutsche
  Forschungsgemeinschaft (DFG, German Research Foundation) within SFB
  1227 (DQ-mat, project A04), SPP 1929 (GiRyd), and under Germany’s
  Excellence Strategy -- EXC-2123 QuantumFrontiers -- 390837967.

\end{acknowledgments}

\appendix

\section{Machine Learning parameters for learning the error correlations \label{app:A}}

In this appendix, we provide details on the neural networks employed
to train the errors associated with gapped quantum phases. As
introduced earlier and as depicted in Fig.~\ref{fig:schematic_rep}, we
consider a DNN with input as the errors, multiple hidden layers and
two outputs. For a given system size, there is a fixed number of
errors which thereby fixes the input nodes to the DNN. Each hidden
layer has approximately half the number of nodes as the previous layer
and the number of hidden layers is increased until the nodes in the
hidden layers are less than 30, with the final output layer having two
nodes.  Having detailed the architecture, we briefly comment on the
tools used to construct the neural networks, as well as the training
parameters employed. We have implemented the above DNN architecture
using FluxML~\cite{Innes2018}, a machine learning library in
Julia. For training purposes, we choose the cross entropy loss
function with the Adam algorithm for optimization.  Further, we set
the learning rate, $\eta$, on the order of $10^{-3}$, and have used
GPUs for the purpose of training. For training and predicting
purposes, we generate around 25,000 error samples for a given
wavefunction.

\section{Error correction circuit used to compute the correction time statistics \label{app:B}}
In this appendix, we briefly review the error correction circuit employed to compute the standard deviation of 
the circuit depth used within the operational definition as shown in Fig.~\ref{fig:toric_hx}. For a given wavefunction, 
we sample the errors as outlined in the main text and for each error sample we perform the error correction 
until all the face excitations are annihilated. To compute the error correction time for 
a given error sample, we follow the procedure from Ref.~\onlinecite{Jamadagni2020} and attach to each error a walker that 
traverses the surrounding in a diamond shaped pattern of increasing Manhattan distance from the error. Whenever two walkers 
start to overlap, their corresponding errors are fused and removed from the system. The circuit depth of a particular error 
sample is then given by the number of steps required until the last errors have been fused.

\bibliographystyle{myaps}
\bibliography{ml_qp}

\end{document}